\documentclass{article}
\usepackage[utf8]{inputenc}
\usepackage{amsmath}
\usepackage{graphicx}
\usepackage{amsfonts}
\usepackage{amssymb}
\usepackage{geometry}
\usepackage{float}
\usepackage{bm}
\usepackage{color}
\usepackage{cite}
\usepackage{subfigure}
\usepackage{geometry}
\usepackage{authblk}
\begin{document}
\title{Thermodynamics of Schwarzschild black hole surrounded by quintessence
in gravity's rainbow}
\author{B. Hamil\thanks{%
hamilbilel@gmail.com (corresponding author)} \\
D\'{e}partement de TC de SNV, Universit\'{e} Hassiba Benbouali, Chlef,
Algeria. \and B. C. L\"{u}tf\"{u}o\u{g}lu\thanks{%
bekircanlutfuoglu@gmail.com } \\
Department of Physics, University of Hradec Kr\'{a}lov\'{e}, \\
Rokitansk\'{e}ho 62, 500 03 Hradec Kr\'{a}lov\'{e}, Czechia.}
\date{}
\maketitle

\begin{abstract}
According to some quantum gravity models, Lorentz invariance can be violated in the Planck energy scale. With this motivation, we analyze the thermal quantities and the stability of Schwarzschild black hole surrounded by quintessence in gravity's rainbow formalism. To do that, we consider the rainbow functions which are motivated by loop quantum gravity and gamma-ray bursts, and we derive Hawking temperature, specific heat, entropy and the equation of state function. We observe that the presence the quintessence matter field and rainbow gravity affect the stability of the black hole.
\end{abstract}

\section{Introduction}
The thermodynamics of black hole (BH) is perhaps one of the most exciting topics in modern cosmology. The original link between BHs and thermodynamics were firstly presented by Bekenstein and Hawking \cite{Bek,Bekenstein,Hawkink}. Half a century ago Bekenstein demonstrated that the entropy, $S$, of BHs should be proportional to their event horizon area, $A$, in Planck units. Soon afterwards, by using the quantum field theory in the context of curved backgrounds, Hawking showed that the BHs can absorb and emit radiation from the event horizon, named as "Hawking radiation", with Hawking temperature 
\begin{eqnarray}
T_{H}=\frac{\hbar \kappa }{2\pi K_{B}}, \label{HawTem}
\end{eqnarray}
where $K_B$ and $\kappa $ correspond to the Boltzmann constant and   the surface gravity of the BHs, respectively. As a well-known fact, the second law of the conventional laws of thermodynamics states that the entropy of a closed system must always increase. In the context of BH thermodynamics, this law emerges similarly to the second law of BH thermodynamics, by  declaring that the BH surface area of the event horizon cannot shrink.

Recent cosmological observations on type Ia supernova \cite{Perlmuter,Zhang}, the cosmic microwave background \cite{Riess,Lamon} and
large scale structure \cite{Tegmark} suggest an accelerating expansion of our Universe. In order to model this observed fact, some form of energy with negative pressure,
called "dark energy", is being integrated to 
the theory of Einstein's general relativity \cite{Sherwin}. Comprehending the origin of negative pressure is one of the greatest travails of modern cosmology, and so dark energy is assumed being one of the primary challenges of cosmology with its convoluted evidences. In literature, we observe two distinct candidate models to realize dark energy: the cosmological constant model \cite{Padmanabhan}, which theoretical and experimental results differ greatly, and the dynamical scalar field models such as quintessence \cite{Carroll}, phantom \cite{Caldwell}, and \textbf{K}-essence \cite{Picon} matters.

The quintessence matter is  modeled by gravity coupled inhomogeneous scalar field in which the equation of states for the quintessence matter is obtained as 
\begin{eqnarray}
P _{q}=\omega _{q}\rho _{q}.
\end{eqnarray}
Here, $\omega _{q}$, named the quintessential state parameter, can take values only in the range of $-1<\omega_{q}<-1/3$. $P _{q}$ and $\rho _{q}$ correspond to the pressure and the density of the quintessence matter, respectively. Two decades ago, Kiselev combined BH physics with the quintessence matter by investigating  static spherically-symmetric solutions of Einstein's field equation \cite{Kiselev}. After his novel results, we observe an increasing number of decent papers in the literature focusing on the Kiselev BHs embedded in quintessence matter \cite{Fernando,Uniyal,Malako,Zheng,Younas,Ghaderi,Jiao,Shahjalal,Eslamzadeh,Saghafi,Ghosh,Dahbi,Wen}.

On the other hand,  the Lorentz symmetry is one of the most essential symmetries of nature, which generates the standard dispersion relation of Einstein's special relativity. However, in the Planck energy scale, $E_{p}\simeq 10^{19}\mathrm{GeV}$, this standard dispersion relation should be altered by incorporating a new fundamental constant.  The deformed energy-momentum dispersion relation comes out in a new theory
named doubly/ deformed special relativity (DSR) \cite{Magueijo, Lee}. The latter theory has two invariants: the speed of light and the Planck energy. An extension of the DSR theory to curved spacetime is known as rainbow gravity (RG), and it refers the geometry of spacetime correlates with the probing particle's energy \cite{Smolin}.

In this manuscript, we intend to examine the thermodynamics of the Schwarzschild BH embedded in quintessence matter in RG formalism, with the particular choices of rainbow functions that are harmonious with outcomes of the loop quantum gravity and gamma-ray bursts. To this end, we construct the main body as follows: In Sec. 2, we review Kiselev's geometry  and derive a general formalism for RG . Then, in Sec. 3, we employ two particular RG functions to examine the stability of BH as well as the RG-corrected Hawking temperature, specific heat, remnant mass, entropy, and equation of state functions. We conclude the manuscript with a brief review of our findings.

\section{General formalism}
One can express the geometry of the Schwarzschild BH embedded in quintessence matter with the metric
\begin{equation}
ds^{2}=-g\left( r\right) dt^{2}+\frac{1}{g\left( r\right) }%
dr^{2}+r^{2}d\Omega ,
\end{equation}
by the following lapse function
\begin{equation}
g\left( r\right) =1-\frac{2M}{r}-\frac{\alpha }{r^{3\omega_q +1}}.
\end{equation}
Here, $M$ and $\alpha $ denote the BH's mass, and the normalization factor, respectively.
In this manuscript, we take into consideration three particular values of quintessence state parameter. The first one, $\omega _{q}=-1$, covers the cosmological constant model while the metric brings to the Schwarzschild de-Sitter BH geometry. The second one, $\omega _{q}=-1/3$, the metric gives simply the Schwarzschild BH with one event horizon value. The third one, $\omega _{q}=-2/3$, the metric permits BH having inner and outer horizons at
\begin{equation}
r_{\pm }=\frac{1\pm \sqrt{1-8\alpha M}}{2\alpha },
\end{equation}
simultaneously. In order to examine the impact of the RG, we substitute  $dt$ and $dx_i$ with $ \frac{dt}{\mathcal{A}\left(
E/E_{p}\right) }$ and $ \frac{dx_{i}}{\mathcal{B}\left(
E/E_{p}\right) }$, respectively. By doing this, we obtain the RG-corrected metric in the form of
\begin{equation}
ds^{2}=-\frac{g\left( r\right) }{\mathcal{A}^{2}\left( E/E_{p}\right) }%
dt^{2}+\frac{1}{\mathcal{B}^{2}\left( E/E_{p}\right) g\left( r\right) }%
dr^{2}+\frac{r^{2}}{\mathcal{B}^{2}\left( E/E_{p}\right) }d\Omega, 
\label{met}
\end{equation}
where RG functions tend to one in low energy scale
\begin{equation}
\lim_{\frac{E}{E_{p}}\rightarrow 0}\mathcal{A}\left( E/E_{p}\right) =\lim_{%
\frac{E}{E_{p}}\rightarrow 0}\mathcal{B}\left( E/E_{p}\right) =1.
\end{equation}
At first, we calculate the RG-corrected Hawking temperature with the given formula \cite{Ahmed}:
\begin{equation}
T_{H}=\frac{\mathcal{B}\left( E/E_{p}\right) }{\mathcal{A}\left(
E/E_{p}\right) }T_{0},  \label{HT}
\end{equation}%
where $T_0$ represents the quintessence matter surrounded BH's Hawking temperature, \cite{Dahbi},  with $K_B=1$.
\begin{equation}
T_{0}=\frac{1}{4\pi r_{H}}\left( 1+\frac{3\alpha \omega }{r_{H}^{3\omega +1}}%
\right). \label{Tem}
\end{equation}%
According to references \cite{Ahmed,Adler,Cavaglia,Medved,Amelino,Ali,Farag}, a lower bound on the energy of a particle that is emitted by BH via Hawking radiation can be obtained from the uncertainty relation between momentum and position. This lower limit is approximately expected to be proportional to the inverse of the radius of the event horizon.
\begin{equation}
E\geq \frac{1}{\Delta x}\sim \frac{1}{r_{H}}.  \label{ap}
\end{equation}%
By using Eqs. (\ref{HT}), (\ref{Tem}) and (\ref{ap}) we obtain RG-corrected Hawking temperature in the form of 
\begin{equation}
T_{H}=\frac{1}{4\pi r_{H}}\left( 1+\frac{3\alpha \omega }{r_{H}^{3\omega +1}}%
\right) \frac{\mathcal{B}\left( 1/E_{p}r_{H}\right) }{\mathcal{A}\left(
1/E_{p}r_{H}\right) }.  \label{FT}
\end{equation}%
We observe that when $\alpha \rightarrow 0$, Eq. \eqref{FT} descends to the Hawking temperature of the RG formalism. Then, we 
investigate the heat capacity function by employing
\begin{equation}
C=\frac{dM}{dT}=\frac{\partial M}{\partial r_{H}}\frac{\partial r_{H}}{%
\partial T}.\label{Cc}
\end{equation}%
In this case,  we get the general form of 
RG-corrected heat capacity  as
\begin{equation}
C=-\frac{2\pi r_{H}^{2}\left( 1+\frac{3\alpha \omega _{q}}{r_{H}^{3\omega +1}%
}\right) }{\left( 1+\frac{3\alpha \omega _{q}\left( 3\omega _{q}+2\right) }{%
r_{H}^{3\omega _{q}+1}}\right) \frac{\mathcal{B}\left( 1/E_{p}r_{H}\right) }{%
\mathcal{A}\left( 1/E_{p}r_{H}\right) }+r_{H}\left( 1+\frac{3\alpha \omega
_{q}}{r_{H}^{3\omega +1}}\right) \frac{d}{dr}\frac{\mathcal{B}\left(
1/E_{p}r_{H}\right) }{\mathcal{A}\left( 1/E_{p}r_{H}\right) }}. \label{CT}
\end{equation}
Next, we ,examine the RG-corrected entropy function. To this end, we consider the first law of BH thermodynamics
\begin{equation}
S=\int \frac{dM}{T}=2\pi \int r_{H}\frac{\mathcal{A}\left(
1/E_{p}r_{H}\right) }{\mathcal{B}\left( 1/E_{p}r_{H}\right) }dr_{H}, \label{ST}
\end{equation}%
which degenerates to $S=\pi r_{H}^{2}$ when $\left( E/E_{p}\right)
\rightarrow 0$. Finally, we intend to derive the equation of state after evaluating the volume. To this end, we use the Legendre transformation and get a  relation between the volume and event horizon
\begin{equation}
V=\left( \frac{\partial M}{\partial P}\right) _{S}=\frac{4\pi }{%
r_{H}^{3\omega _{q}}}\rightarrow r_{H}=\left( \frac{4\pi }{V}\right) ^{\frac{%
1}{3\omega _{q}}}, \label{VT}
\end{equation}
then, via the relation of  pressure and the normalization parameter 
\begin{equation}
P=-\frac{\alpha }{8\pi }.
\end{equation}
we set the RG-corrected equation of state of the BH as%
\begin{equation}
P=\frac{1}{24\pi \omega }\left( \frac{4\pi }{V}\right) ^{\frac{3\omega +1}{%
3\omega _{q}}}\left\{ 1-4\pi \left( \frac{4\pi }{V}\right) ^{\frac{1}{%
3\omega _{q}}}T_{H}\frac{\mathcal{A}\left( 1/E_{p}r_{H}\right) }{\mathcal{B}%
\left( 1/E_{p}r_{H}\right) }\right\} . \label{PT}
\end{equation}%
We observe that the thermal quantities and stability of the BH depend on the choice of the RG functions.

\section{Applications}
In this section, we take into consideration two specific choices of RG functions and  examine the BH thermodynamic functions.

\subsection{Loop quantum gravity}
Our first choice is harmonious with outcomes of loop quantum gravity and $\kappa $- Minkowski noncommutative spacetime \cite{Ellis,Cam,Hooft,Samuel}
\begin{equation}
\mathcal{A}=1, \quad\quad \mathcal{B}=\sqrt{1-\eta \frac{E^{2}}{E_{p}^{2}}}. \label{r}
\end{equation}%
In this case, Eq. \eqref{FT} gives
\begin{equation}
T_{H}=\frac{1}{4\pi r_{H}}\left( 1+\frac{3\alpha \omega _{q}}{r_{H}^{3\omega
_{q}+1}}\right) \sqrt{1-\frac{\eta }{E_{p}^{2}r_{H}^{2}}}.  \label{L1}
\end{equation}%
Obviously, if one sets $\eta =0$, then RG effect disappears and  Eq. \eqref{L1} gives the Hawking temperature of Schwarzschild black hole surrounded by quintessence matter. Now, we analyze the RG-corrected Hawking temperature function for three distinct quintessential state parameter values: 
\begin{itemize}
\item In the $\omega _{q}=-2/3$ case,  Eq.(\ref{L1}) becomes%
\begin{equation}
T_{H}=\frac{1-2\alpha r_{H}}{4\pi r_{H}}\sqrt{1-\frac{\eta }{%
E_{p}^{2}r_{H}^{2}}}.
\end{equation}%
We see that to obtain a physical temperature, we must impose  the following conditions $\left( 1-2\alpha r_{H}\right)
\geq 0$, and $1-\frac{\eta }{E_{p}^{2}r_{H}^{2}}\geq 0$, which dictate upper and lower bound values on event horizon radius $\frac{1}{2\alpha }\geq r_{H}\geq 
\frac{\sqrt{\eta }}{E_{p}}$.

\item In the $\omega _{q}=-1$ case,  Eq. (\ref{L1}) simplifies to%
\begin{equation}
T_{H}=\frac{1-3\alpha r_{H}^{2}}{4\pi r_{H}}\sqrt{1-\frac{\eta }{%
E_{p}^{2}r_{H}^{2}}},  \label{TT}
\end{equation}%
which states that event horizon is bounded as $\frac{1}{\sqrt{3\alpha }}\geq r_{H}\geq \frac{\sqrt{\eta }}{E_{p}}.$

\item In the $\omega _{q}=-1/3$ case, Eq.(\ref{L1}) reduces to
\begin{equation}
T_{H}=\frac{1-\alpha }{4\pi r_{H}}\sqrt{1-\frac{\eta }{E_{p}^{2}r_{H}^{2}}}, 
\end{equation}%
which enforces a bound only from below on event horizon radius as
$r_{H}\geq \frac{\sqrt{\eta }}{E_{p}}$.
\end{itemize}

\begin{figure*}[htb]
\resizebox{\linewidth}{!}{\includegraphics{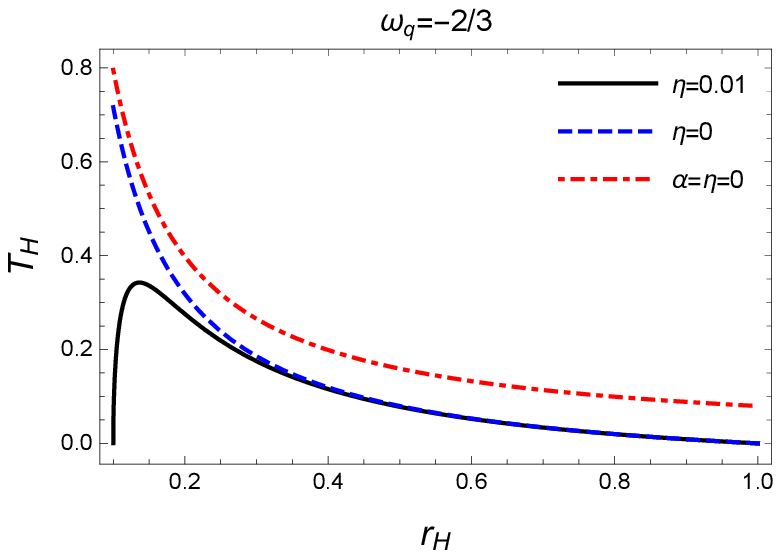},%
\includegraphics{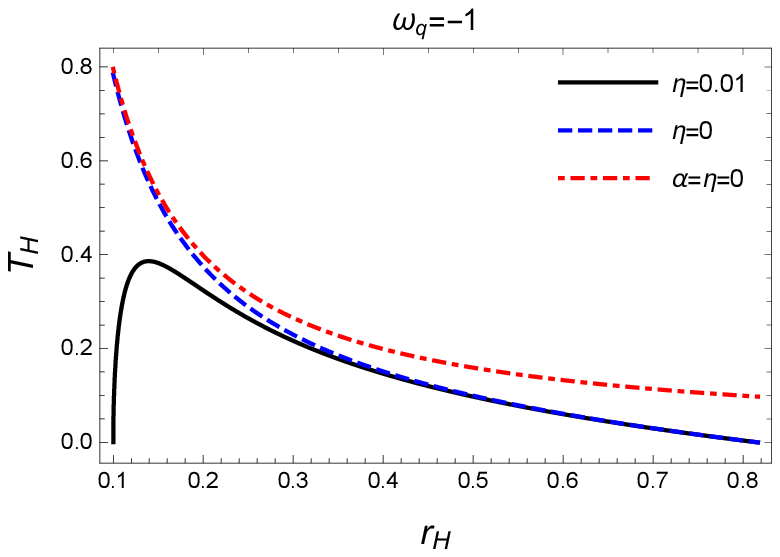},\includegraphics{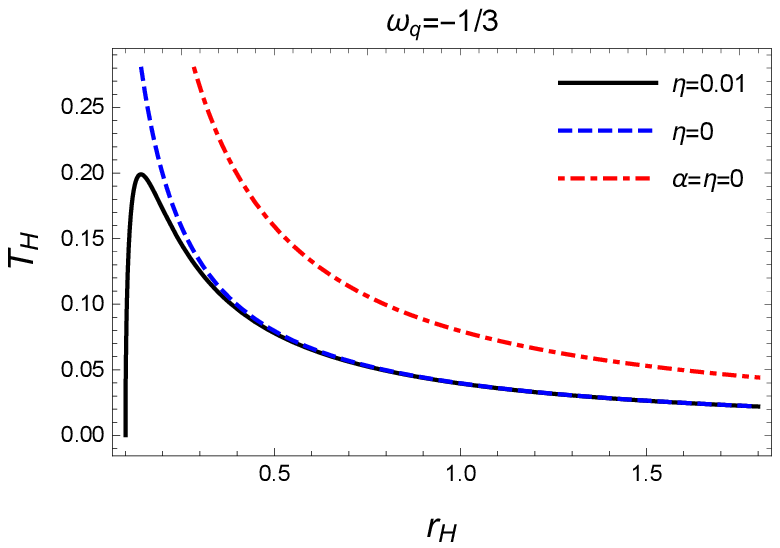}}
\caption{Temperature versus horizon radius for $\alpha=0.5$ and $E_{p}=1$.} \label{Fig1}
\end{figure*}

In Fig. \ref{Fig1}, we depict the variation of the RG-corrected
Hawking temperature with respect to the horizon radius for the distinct values
of $\eta $, $\omega _{q}$ and $\alpha $. 

We observe that quantum corrections, due to the RG and quintessence matter field, have significant effects on the BH temperature only at small event horizon values. In Table 1 we present  numerical results of the maximum temperature for different values of gravity's rainbow parameter. In congruence with Table 1, as $\eta $ becomes smaller the critical radius $r_{c}$ gets smaller too, but the temperature peak becomes more significant.
\begin{table}[htbp]
\centering
\begin{tabular}{|c|c|c|c|c|c|c|}
\hline
$\alpha=0.5$ & \multicolumn{2}{|c}{$\omega_{q}=-2/3$} & \multicolumn{2}{|c}{$%
\omega_{q}=-1$} & \multicolumn{2}{|c|}{$\omega_{q}=-1/3$}  \\ \hline\hline
$\eta$ & $r_{c}$ & $T_{max}$ & $r_{c}$ & $T_{max}$ & $r_{c}$ & $T_{max}$ \\ 
\hline
$10^{-1}$ & 0.4 & 0.0730966 & 0.375035 & 0.0659396 & 0.447212 & 0.0629115 \\ 
$10^{-2}$ & 0.13651 & 0.342649 & 0.13757 & 0.374691 & 0.141421 & 0.198944 \\ 
$10^{-3}$ & 0.0442242 & 1.20228 & 0.0445886 & 1.2507 & 0.0447212 & 0.629115
\\ 
$10^{-4}$ & 0.0140922 & 3.9227 & 0.0141379 & 3.97649 & 0.0141421 & 1.98944
\\ 
$10^{-5}$ & 0.00446714 & 12.5261 & 0.004472 & 12.5815 & 0.00447212 & 6.29115
\\ \hline\hline
\end{tabular}%
\caption{Correlation between  critical horizon radius and maximum temperature for different values of $\eta $.}
\end{table}

Next, we examine the
thermal stability of the Schwarzschild black hole surrounded by
quintessence in gravity's rainbow. We know that a black hole is considered stable when its heat capacity function has positive values and unstable when it has negative values. Recalling the definition given in Eq. \eqref{Cc}, we can determine the stability via the sign of $\left( \frac{\partial T_{H}}{\partial r_{H}}\right) ^{-1}$. When the temperature get its maximum value at $r_H=r_{c}$, then the slope of the
temperature curve becomes zero $\left. \frac{\partial T_{H}}{\partial r_{H}}
=0\right\vert _{r_{H}=r_{c}}$, and  for this special value the heat capacity
becomes singular. Thus, the thermal quantity is separated into two parts: stable early stage and unstable late stage.

\begin{figure*}[htb]
\resizebox{\linewidth}{!}{\includegraphics{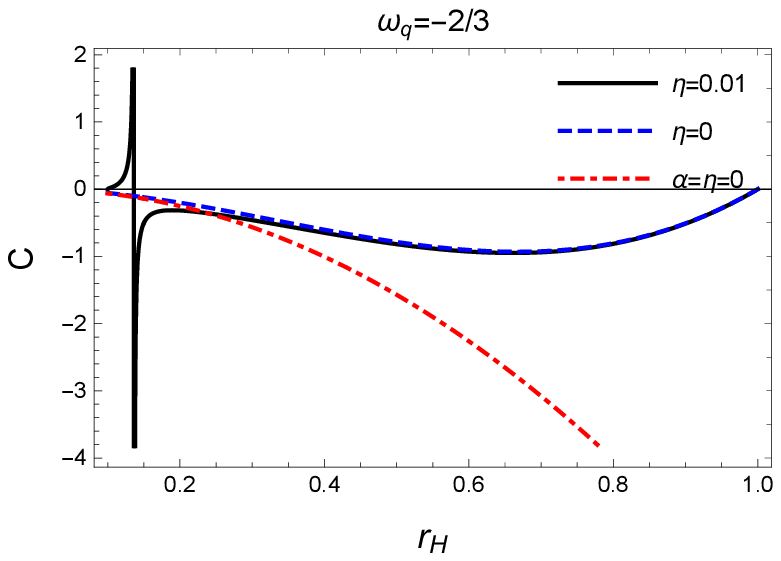},%
\includegraphics{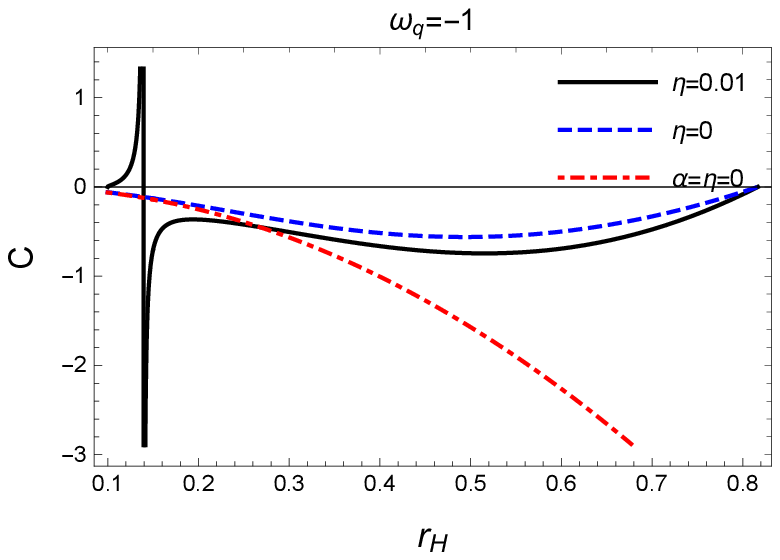},\includegraphics{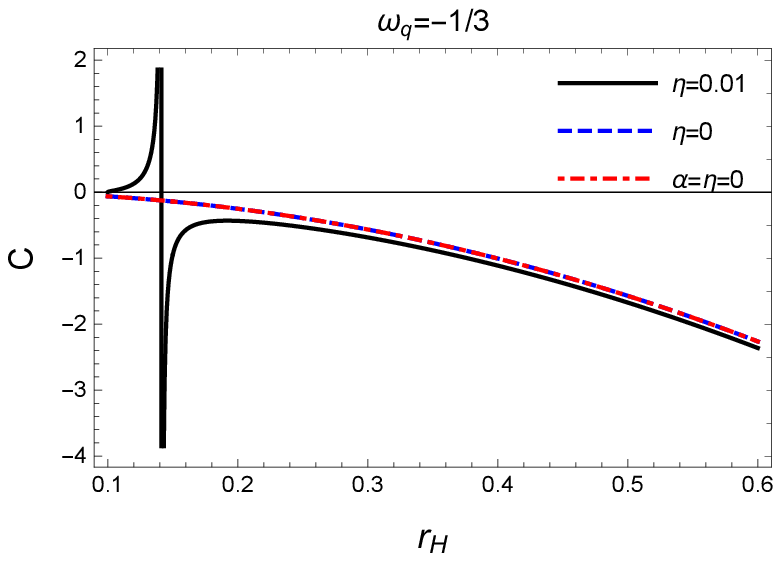}}
\caption{Heat capacity versus horizon radius for  $\alpha=0.5$ and $E_{p}=1$.} \label{Fig2}
\end{figure*}

Now we employ Eq. \eqref{CT} and Eq. \eqref{r} to express the GR-corrected specific heat function 
\begin{equation}
C=-\frac{2\pi r_{H}^{2}\left( 1+\frac{3\alpha \omega _{q}}{r^{3\omega +1}}%
\right) \sqrt{1-\frac{\eta }{E_{p}^{2}r_{H}^{2}}}}{\left( 1+\frac{3\alpha
\omega _{q}\left( 3\omega _{q}+2\right) }{r_{H}^{3\omega _{q}+1}}\right) -%
\frac{\eta }{E_{p}^{2}r_{H}^{2}}\left( 2+\frac{3\alpha \omega _{q}\left(
3\omega _{q}+3\right) }{r_{H}^{3\omega _{q}+1}}\right) }.  \label{C}
\end{equation}
It is worthwhile note that when the heat capacity becomes zero a black hole is not exchanging any heat with its surrounding. In this context, Eq. \eqref{C} implies presence of remnant mass when the horizon radius takes the following values 
\begin{equation}
r_{rem}=\left( -3\alpha \omega _{q}\right) ^{\frac{1}{3\omega _{q}+1}}\text{
\ and \ }r_{rem}=\frac{\sqrt{\eta }}{E_{p}}\text{ }.
\end{equation}%
We observe that the remnant horizon radius depends on the quintessence parameters $\left(\alpha ,\omega _{q}\right) $ and rainbow gravity parameters $\eta $. According to the remnant radii values, we obtain the corresponding remnant mass values as
\begin{equation}
M_{rem}=\frac{\left( -3\alpha \omega _{q}\right) ^{\frac{1}{3\omega _{q}+1}}%
}{2}-\frac{\alpha }{2\left( -3\alpha \omega _{q}\right) ^{\frac{3\omega _{q}%
}{3\omega _{q}+1}}}\text{ \ and }M_{rem}=\frac{\sqrt{\eta }}{2E_{p}}-\frac{%
\alpha }{2}\left( \frac{E_{p}}{\sqrt{\eta }}\right) ^{3\omega _{q}}.
\label{M}
\end{equation}
Now, we take into consideration the following particular cases and express the GR-corrected specific heat function with remnant mass values.
\begin{itemize}
\item In the $\omega _{q}=-2/3$ case: 
\begin{equation}
C=-\frac{2\pi r_{H}^{2}\left( 1-2\alpha r_{H}\right) \sqrt{1-\frac{\eta }{%
E_{p}^{2}r_{H}^{2}}}}{1-\frac{2\eta }{E_{p}^{2}r_{H}^{2}}\left( 1-\alpha
r_{H}\right) }.
\end{equation}%
\begin{equation}
M_{rem}=\frac{1}{8\alpha },\text{ \ and }M_{rem}=\frac{\sqrt{\eta }}{2E_{p}}%
\left( 1-\frac{\alpha \sqrt{\eta }}{E_{p}}\right) .
\end{equation}

\item In the $\omega _{q}=-1$ case:
\begin{equation}
C=-\frac{2\pi r_{H}^{2}\left( 1-3\alpha r_{H}^{2}\right) \sqrt{1-\frac{\eta 
}{E_{p}^{2}r_{H}^{2}}}}{\left( 1+3\alpha r_{H}^{2}\right) -\frac{2\eta }{%
E_{p}^{2}r_{H}^{2}}}.
\end{equation}%
\begin{equation}
M_{rem}=\frac{1}{3\sqrt{3\alpha }},\text{ \ \ and }M_{rem}=\frac{\sqrt{\eta }%
}{2E_{p}}\left( 1-\frac{\alpha \eta }{E_{p}^{2}}\right) .
\end{equation}

\item In the  $\omega _{q}=-1/3$ case:
\begin{equation}
C=-\frac{2\pi r_{H}^{2}\sqrt{1-\frac{\eta }{E_{p}^{2}r_{H}^{2}}}}{1-\frac{%
2\eta }{E_{p}^{2}r_{H}^{2}}}.
\end{equation}%
\begin{equation}
M_{rem}=\frac{\sqrt{\eta }\left( 1-\alpha \right) }{2E_{p}}.
\end{equation}
\end{itemize}
Unfortunately, it is not easy to perform an analytical analysis of the RG-corrected heat capacity function. Therefore, we prefer to demonstrate its behavior versus the event horizon regarding distinct values of $\eta $, $\omega _{q}$ and $\alpha $ in  Fig. \ref{Fig2}.

We observe that when $\eta \neq 0$, the heat capacity function presents a discontinuous character at $r_{H}=r_{c}$, and changes it sign nearby this critical radius. To put it more clearly, it is positive for $r_{c}>r_{H}$, thus, points out a stable BH, and it is negative for $r_{c}<r_{H}$, so it identifies an unstable BH in the context of thermodynamics. 

\begin{figure*}[tbh]
\resizebox{\linewidth}{!}{\includegraphics{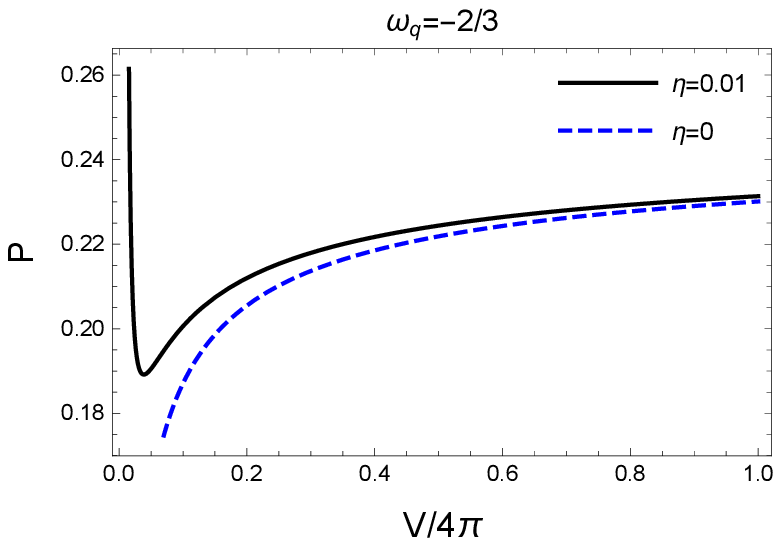},%
\includegraphics{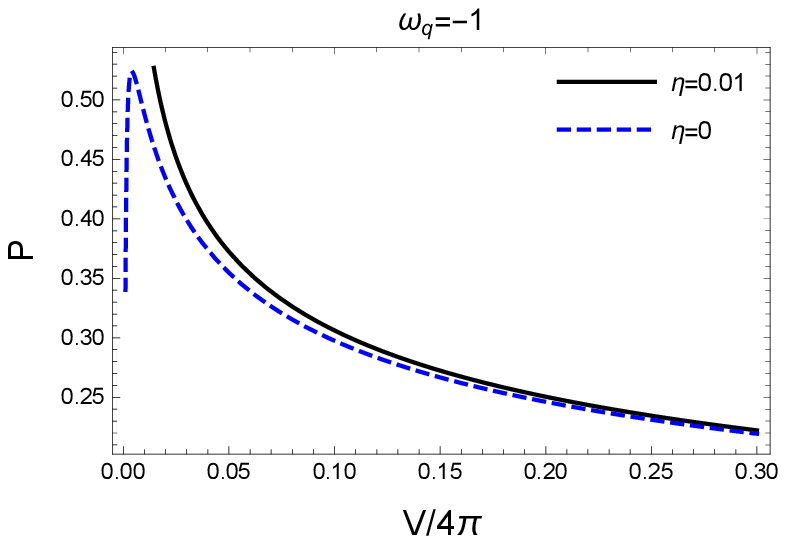},\includegraphics{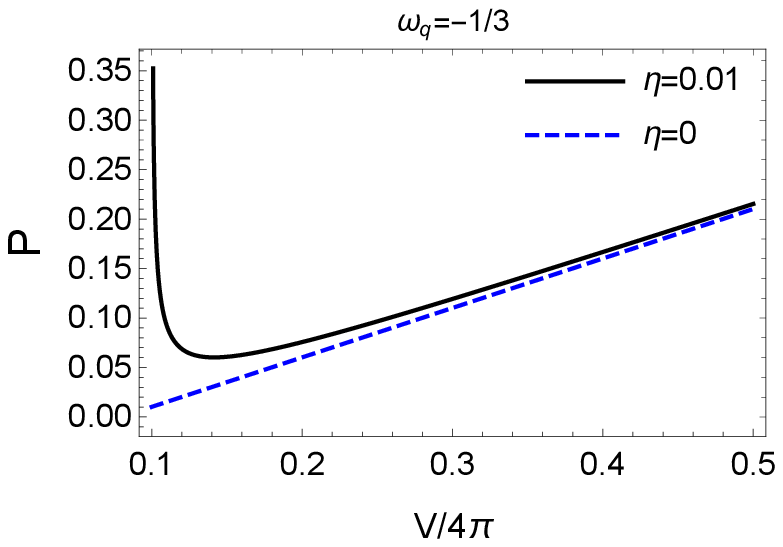}}
\caption{Equation of state versus reduced volume for $T_H=1$, $\alpha=0.5$ and $E_{p}=1$.} \label{Fig3}
\end{figure*}

Next, we obtain the RG-corrected entropy function via Eq. \eqref{ST}. We find
\begin{equation}
S=\pi r_{H}^{2}\sqrt{1-\frac{\eta }{E_{p}^{2}r_{H}^{2}}}+\frac{\pi \eta }{%
E_{p}^{2}}\log \left[ r_{H}+\sqrt{r_{H}^{2}-\frac{\eta }{E_{p}^{2}}}\right] .
\end{equation}%
It is obvious that, one can recover the area law of the entropy by setting $\eta =0$. Then, we obtain the RG-corrected equation of state function by using Eqs. (\ref{PT}) and (\ref{r}).
\begin{equation}
P=\frac{1}{24\pi \omega _{q}}\left( \frac{4\pi }{V}\right) ^{\frac{3\omega
_{q}+1}{3\omega _{q}}}\left \{ 1-\frac{4\pi \left( \frac{4\pi }{V}\right) ^{%
\frac{1}{3\omega _{q}}}}{\sqrt{1-\frac{\eta }{E_{p}^{2}}\left( \frac{V}{4\pi 
}\right) ^{\frac{2}{3\omega _{q}}}}}T_{H}\right \} .\end{equation}
In Fig. \ref{Fig3}, we plot RG-corrected $P-V$ isotherm versus reduced volume by considering distinct
values of $\eta $, $\omega _{q}$ and $\alpha $.
We observe that RG corrections alter the equation of state isotherms only in relatively small volumes.

\subsection{Gamma - ray bursts}
\begin{figure*}[htb]
\resizebox{\linewidth}{!}{\includegraphics{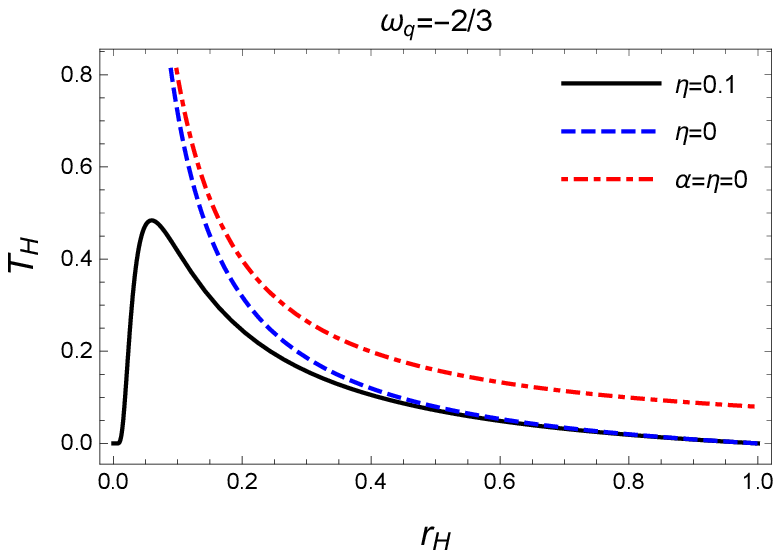},%
\includegraphics{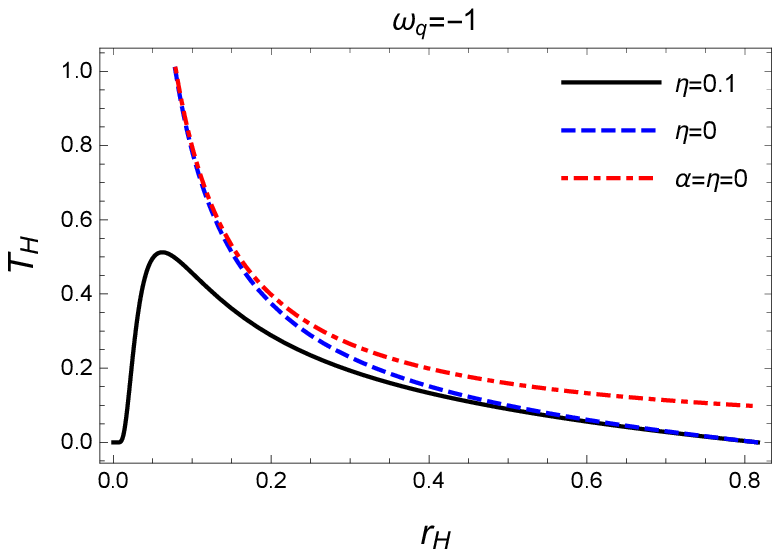},\includegraphics{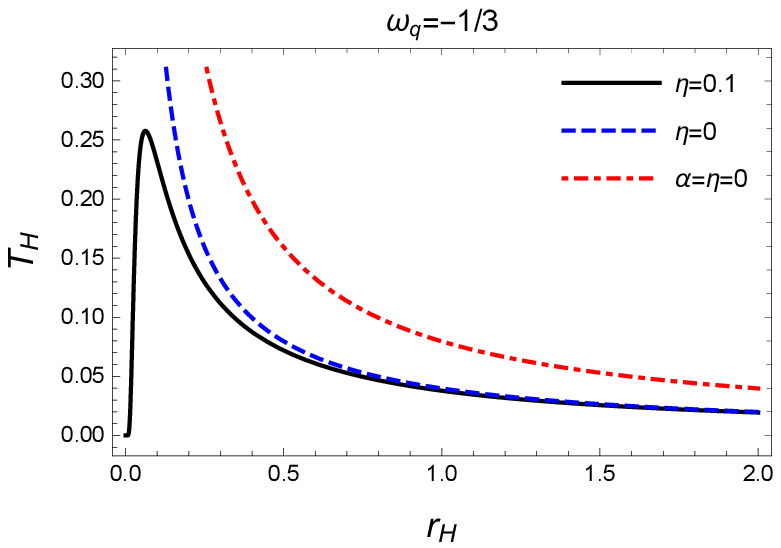}}
\caption{Temperature versus horizon radius for $\alpha=0.5$ and $E_{p}=1$.} \label{Fig4}
\end{figure*}

Our second choice is harmonious with outcomes of 
he hard spectra of gamma-ray bursts at cosmological distance \cite{Nanopoulos}.
\begin{equation}
\mathcal{A}=\frac{e^{\eta \frac{E}{E_{p}}}-1}{\eta \frac{E}{Ep}}, \quad\quad \mathcal{B}=1.  \label{rai}
\end{equation}%
In this case, Eq. \eqref{FT} gives the RG-corrected Hawking temperature in the form of
\begin{equation}
T_{H}=\frac{\eta }{4\pi E_{p}r_{H}^{2}}\frac{1+\frac{3\alpha \omega _{q}}{%
r_{H}^{3\omega +1}}}{e^{\frac{\eta }{E_p r_{H}}}-1}.  \label{G}
\end{equation}%
By considering real-valued temperature we obtain bound values on event horizon which also depends on quintessence matter parameters, $\omega_q$ and $\alpha$. To clarify this, we investigate the following particular scenarios:
\begin{itemize}
\item In the $\omega _{q}=-2/3$ case,  Eq. (\ref{G}) becomes%
\begin{equation}
T_{H}= \frac{\eta }{4\pi E_{p}r_{H}^{2}}\frac{1-2\alpha r_{H}}{e^{\frac{\eta }{E_p r_{H}}}-1},
\end{equation}%
with $r_{H}\leq \frac{1}{2\alpha }$.

\item In the $\omega _{q}=-1$ case,  Eq. (\ref{G}) simplifies to%
\begin{equation}
T_{H}=\frac{\eta }{4\pi E_{p}r_{H}^{2}}\frac{1-3\alpha r_{H}^{2}}{e^{\frac{\eta }{E_p r_{H}}}-1},
\end{equation}%
with  $r_{H}\leq \frac{1}{\sqrt{3\alpha }}$.

\item In the $\omega _{q}=-1/3$ case, Eq. (\ref{G}) reduces to
\begin{equation}
T_{H}=\frac{\eta }{4\pi E_{p}r_{H}^{2}}\frac{1-\alpha}{e^{\frac{\eta }{E_p r_{H}}}-1},
\end{equation}%
does not dictate any bound on horizon.
\end{itemize}
In Fig. \ref{Fig4} we display RG-corrected Hawking temperature versus horizon for different values of $\left( \omega _{q},\alpha ,\eta \right) $. 

Through the figure, we see that for $\eta=0$ temperature decreases while the horizon radius increases. Setting a non zero $\eta$ parameter changes the behavior and the Hawking temperature increases and reaches a peak, where the evaporation of black hole is maximum before the decrease. 

Then, in this scenario we derive the RG-corrected heat capacity. We find
\begin{equation}
C=-\frac{\pi E_{p}r_{H}^{3}}{\eta }\frac{\left( 1+\frac{3\alpha \omega _{q}}{%
r_{H}^{3\omega +1}}\right) \left( e^{\frac{\eta }{E_{p}r_{H}}}-1\right) }{1+%
\frac{3\alpha \omega _{q}\left( 3\omega _{q}+3\right) }{2r_{H}^{3\omega
_{q}+1}}-\frac{\eta }{2E_{P}r_{H}}\left( 1+\frac{3\alpha \omega _{q}}{%
r_{H}^{3\omega +1}}\right) \frac{e^{\frac{\eta }{E_{p}r_{H}}}}{\left( e^{%
\frac{\eta }{E_{p}r_{H}}}-1\right) }}. \label{Cn}
\end{equation}%
Eq. \eqref{Cn} goes to zero at $r_{rem}=\left(
-3\alpha \omega _{q}\right) ^{\frac{1}{3\omega _{q}+1}}$ and predicts a remnant mass.

\begin{figure*}[htb]
\resizebox{\linewidth}{!}{\includegraphics{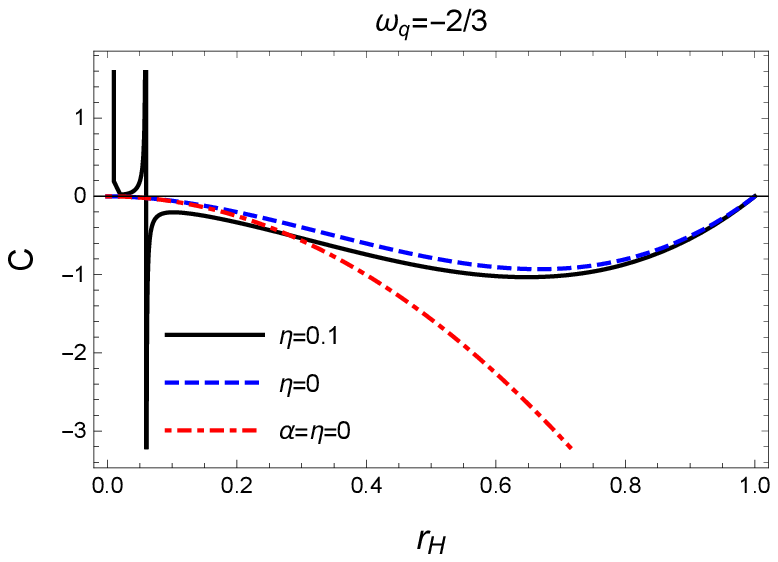},%
\includegraphics{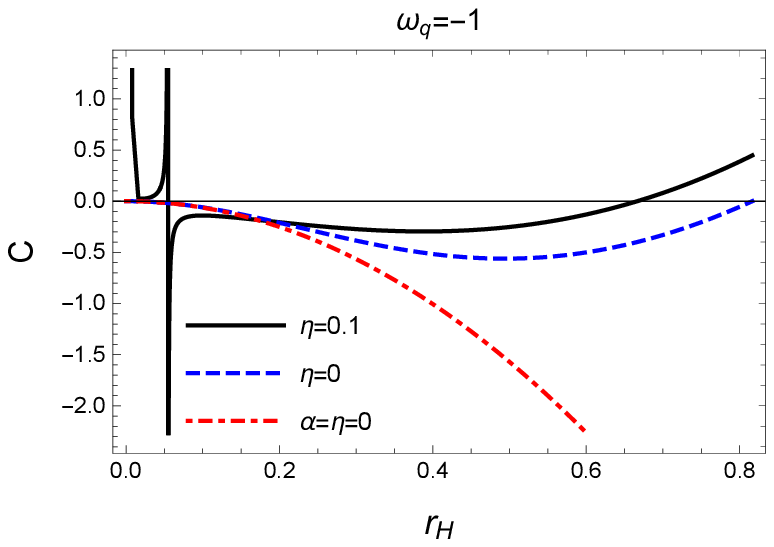},\includegraphics{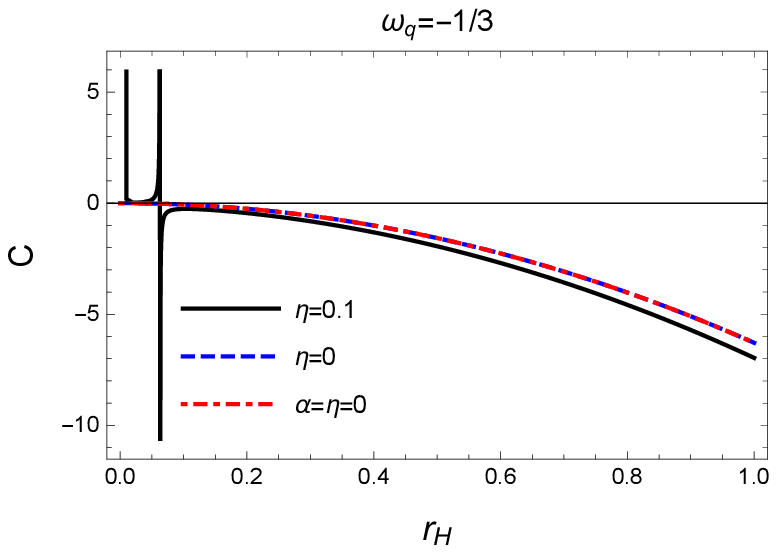}}
\caption{Heat capacity versus horizon radius for  $\alpha=0.5$ and $E_{p}=1$.} \label{Fig5}
\end{figure*}

\begin{itemize}
\item In the $\omega _{q}=-2/3$ case: 
\begin{equation}
C=-\frac{\pi E_{p}r_{H}^{3}}{\eta }\frac{\left( 1-2\alpha r_{H}\right) \left( e^{\frac{\eta }{E_{p}r_{H}}}-1\right) }{1-\alpha r_H%
-\frac{\eta }{2E_{P}r_{H}}\left( 1-2\alpha r_{H}\right) \frac{e^{\frac{\eta }{E_{p}r_{H}}}}{\left( e^{%
\frac{\eta }{E_{p}r_{H}}}-1\right) }},
\end{equation}
with $M_{rem}=\frac{1}{8 \alpha}$.

\item In the $\omega _{q}=-1$ case:
\begin{equation}
C=-\frac{\pi E_{p}r_{H}^{3}}{\eta }\frac{\left( 1-3\alpha r_{H}^2 \right) \left( e^{\frac{\eta }{E_{p}r_{H}}}-1\right) }{1-\frac{\eta }{2E_{P}r_{H}}\left( 1-3\alpha r_{H}^2 \right) \frac{e^{\frac{\eta }{E_{p}r_{H}}}}{\left( e^{%
\frac{\eta }{E_{p}r_{H}}}-1\right) }},
\end{equation}%
with $M_{rem}=\frac{1}{3\sqrt{3\alpha }}$.

\item In the  $\omega _{q}=-1/3$ case:
\begin{equation}
C=-\frac{\pi E_{p}r_{H}^{3}}{\eta }\frac{\left( 1-\alpha \right) \left( e^{\frac{\eta }{E_{p}r_{H}}}-1\right) }{1-\frac{\eta }{2E_{P}r_{H}}\left( 1- \alpha \right) \frac{e^{\frac{\eta }{E_{p}r_{H}}}}{\left( e^{%
\frac{\eta }{E_{p}r_{H}}}-1\right) }}.
\end{equation}%
In this case, there is no remnant mass. 
\end{itemize}
In Fig. \ref{Fig5} we show the RG-corrected heat capacity functions of the scenario.

It is interesting to note that a
critical horizon radius appear in each cases in which the sign of the heat capacity function changes. Therefore, we conclude that this scenario can allow a phase transition.

\begin{figure*}[htb]
\resizebox{\linewidth}{!}{\includegraphics{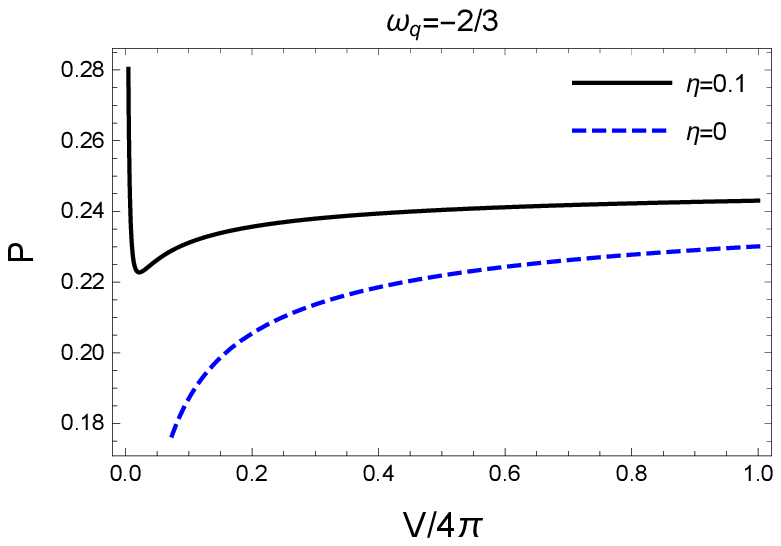},%
\includegraphics{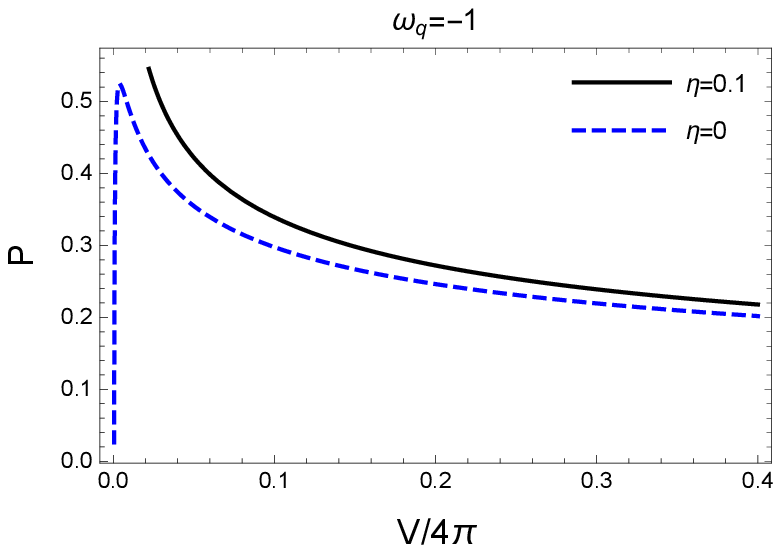},\includegraphics{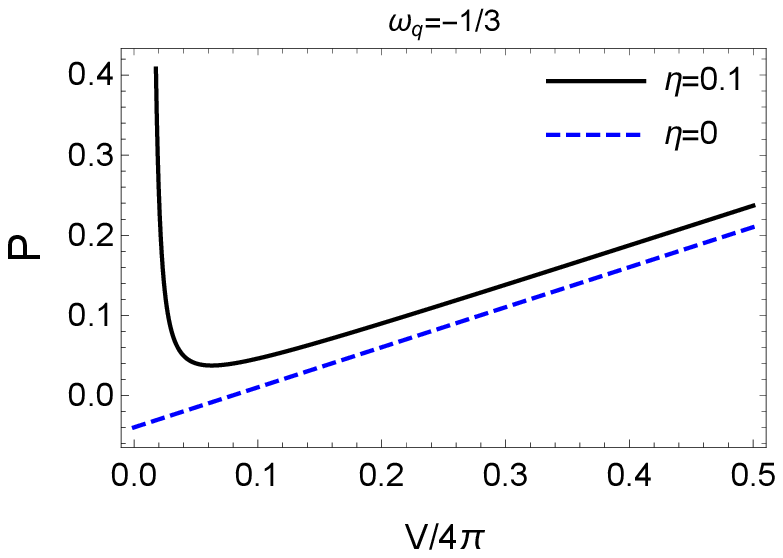}}
\caption{Equation of state versus reduced volume for $T_H=1$, $\alpha=0.5$ and $E_{p}=1$.} \label{Fig6}
\end{figure*}

Then, we explore the RG-corrected entropy function. By using  Eq. \eqref{ST} within this scenario, we get
\begin{equation}
S=\frac{2\pi E_{p}}{3\eta }\left[ -r_{H}^{3}+r_{H}^{3}e^{\frac{\eta }{%
E_{p}r_{H}}}\left( 1+\frac{\eta }{2E_{p}r_{H}}+\frac{\eta ^{2}}{%
2E_{p}^{2}r_{H}^{2}}\right) -\frac{\eta ^{3}}{2E_{p}^{3}}\operatorname{Ei}\left( 
\frac{\eta }{E_{p}r_{H}}\right) \right] .
\end{equation}%
Here, $\operatorname{Ei}$ is the integral exponential function. We see that the RG-corrected entropy function depends only on the horizon radius and rainbow
gravity parameters. Finally, we derive the RG-corrected equation of state function. We find
\begin{equation}
P=\frac{1}{24\pi \omega _{q}}\left( \frac{4\pi }{V}\right) ^{\frac{3\omega +1%
}{3\omega _{q}}}\left[ 1-\frac{4\pi E_{p}}{\eta }\left( \frac{4\pi }{V}%
\right) ^{\frac{2}{3\omega _{q}}}\left( e^{\frac{\eta }{E_{p}}\left( \frac{V%
}{4\pi }\right) ^{\frac{1}{3\omega _{q}}}}-1\right) T_{H}\right] . \label{PV2}
\end{equation}
In Fig. \ref{Fig6}, we depict Eq. \eqref{PV2} versus the reduced volume. 

We see that in all scenarios  P-V isotherms behave similar in greater volume size, while significant difference are observable only in smaller volume size.

\section{Conclusion}
In this manuscript, we study the thermodynamics of the Schwarzschild black hole surrounded by quintessence matter in rainbow gravity formalism. After obtaining a general form for the thermal quantities, we choose two particular rainbow gravity function sets and derive modified Hawking temperature initially. Since the derived function is associated to the quintessence parameters, we consider three particular quintessence state parameters and obtain bounds on the horizon radii. We observe that for the rainbow gravity functions that are related to loop quantum gravity, upper and lower bounds emerge in terms of quintessence and rainbow gravity parameters, respectively. In the second choice of rainbow gravity functions, which are correlated to the gamma bursts of cosmology, we observe that a lower bound limit does not exist.  Then, we examine the specific heat function. We find that within the first choice of rainbow gravity functions, two remnants can occur. However, according to the second choice only one is allowed. Then, we calculate the entropy function. As one can expect, we find a logarithmic correction term in loop quantum gravity approach. Finally, we obtain the equation of state functions and display P-V isotherms to determine the effects of rainbow gravity formalism. After all, we conclude that stability of black hole is related to the choice of rainbow gravity scenarios.

\end{document}